\documentclass[hyper,12pt,paper]{article}
\pdfoutput=1 

\usepackage{jheppub} 

\usepackage[T1]{fontenc} 
\usepackage{tabu}
\usepackage{graphicx,amsmath,amssymb,multirow,array,bm,mathrsfs}
\usepackage{upgreek}

\def\vev#1{\langle\, #1 \, \rangle} 
\def\ket#1{\mid \! #1\rangle} 
\def\bra#1{\langle \, #1 \! \mid}

\def\ptr#1{{{#1}^\Gamma}}
\def\domd{\lozenge_{\cal A}}  
\def\Eneg{{\mathscr E}}
\def\Neg{{\mathscr N}}

\definecolor{rust}{rgb}{0.8,0.2,0.2}

\subheader{}
\title{Comments on Entanglement Negativity in Holographic Field Theories}

\author{Mukund Rangamani,}
\author{\! Massimiliano Rota}
\affiliation{ Centre for Particle Theory \& Department of Mathematical Sciences,\\
                     Science Laboratories, South Road, Durham DH1 3LE, UK.}
\emailAdd{mukund.rangamani@durham.ac.uk}
\emailAdd{massimiliano.rota@durham.ac.uk}

\abstract{
We explore entanglement negativity, a measure of the distillable entanglement contained in a quantum state, in relativistic field theories in various dimensions. We first give a general overview of negativity and its properties and then explain a well known result relating (logarithmic) negativity of pure quantum states to the Renyi entropy (at index $1/2$), by exploiting the simple features of entanglement in thermal  states. In particular,  we show that the negativity of the thermofield double state is given by the free energy difference of the system at temperature $T$ and $2\,T$ respectively. We then use this result to compute the negativity in the vacuum state of conformal field theories in various dimensions, utilizing results that have been derived for free and holographic CFTs in the literature. We also comment upon general lessons to be learnt about negativity in holographic field theories.
}

\begin{document} 
	\begin{flushright} \small{DCPT-14/27} \end{flushright}
	
\maketitle
\flushbottom

\section{Introduction}
\label{sec:intro}

Quantum mechanics, as is well appreciated, is characterized by an important feature, entanglement. While the colloquial usage of the word ``entanglement'' often simply refers to presence of correlations which could simply be of classical nature, nature of quantum entanglement transcends this interpretation. A natural question is to segregate and quantify in a given quantum state the genuinely quantum parts of entanglement from those that are inherited from underlying classical correlations.

One way to proceed would be to use the intuition garnered from Einstein-Podolsky-Rosen 
(EPR) like entangled states, which are non-product (pure) states in the quantum Hilbert space. One of the characteristic hallmarks of these states as elucidated by Bell \cite{Bell:1964kc} is that they fail to satisfy the Bell inequality (and hence its generalization, the CHSH inequalities). We now understand quite well that this means that the entanglement inherent in the EPR state is a genuine quantum aspect and relatedly that one cannot invoke some local hidden variable (LHV) to describe the quantum state. A-priori one might have thought that the Bell/CHSH inequalities provide a complete characterization of the nature of entanglement. 

While for pure states this is true, the state of affairs is much less clear in case of mixed states. Consider a bipartite system in a state $\rho$ with two Hilbert spaces which we will refer to as the left and right Hilbert space, ${\cal H}_L$ and ${\cal H}_R$ respectively. Such a state is called {\em separable} if it can be written as
\begin{equation}
\rho=\sum_{i}\,p_i\,\rho_i^R\otimes\rho_i^L\,,
 \hspace{1cm} \sum_ip_i=1 \,,\hspace{1cm} p_i\geq0
\,,
\label{}
\end{equation}	
otherwise it is called {\em entangled}. Physically this definition attempts to encode the fact that separable states are classically correlated as they can be produced using only local operations and classical communication (LOCC).\footnote{ LOCC for two parties consists of steps in each of which any party is allowed to perform local measurements and communicate the outcome to the other using classical channels.} In particular, it introduces a distinction between the correlations that are classical and those that ought to be considered quantum. 

In analogy with pure states above, one would then be inclined to think that even in the case of mixed states any entangled state violates some Bell inequality. Surprisingly this is not true, as demonstrated by Werner in \cite{Werner:1989aa}, where  mixed entangled states that can nevertheless be described by a LHV model were constructed. In some sense,  despite manifesting some quantum correlation, these states  ought to be viewed as {\em local} as they are not in tension with the notion of local realism.\footnote{ For a discussion on the properties of Werner states in the context of teleportation see e.g., \cite{Popescu:1994aa}.} Furthermore, if we have access to several copies of the state then it is sometimes possible using only LOCC to {\em distill}  a new state that violates some Bell inequality \cite{Popescu:1995aa} (see \cite{Bennett:1995ra} for details on distillation). One might then be led to the intuition that this process should be achievable starting from any mixed state; therefore the only states that always satisfy all Bell inequalities are the separable ones. Unfortunately, even this intuition fails; to put it mildly the boundary between classicality and quantumness is rather fuzzy with no clear demarcation. The main lesson we wish to emphasize is one ought to distinguish different notions of entanglement in the quantum realm.

Because of the intricate nature of entanglement for mixed states, several measures of entanglement have been proposed. The concept of distillation for example can be used to define the {\em distillable entanglement} as a measure of how much pure entanglement it is possible to extract from some state using only LOCC. On the other hand the {\em entanglement of formation} quantifies the amount of pure entanglement required to create a given state.\footnote{ These measures must be interpreted in an asymptotic sense. They give extremal rates achievable when one has many copies of the state $\rho$.} In case of pure states these measures are equal and agree with entanglement entropy (for a comprehensive review on entanglement measures see \cite{Plenio:2007aa}). Unfortunately the drawback is that these measures cannot be computed because they are given by variational expressions over possible LOCC protocols. A more pragmatic approach is to therefore consider a quantity that is computable \cite{Vidal:2002zz} -- this leads us to the consideration of  {\em entanglement negativity} which will form the focus of the present investigation. Heuristically, the concept uses the spectral data of the density matrix\footnote{ We actually need the spectral data of an auxiliary matrix constructed from the density matrix; we will be more precise below.} (sometimes called entanglement spectrum) to ascertain the amount of entanglement inherent in the mixed state (cf., \S\ref{sec:negative} for a precise definition) .

While the above discussion has been firmly rooted in the realm of quantum mechanics, one expects that many of these issues generalize to relativistic quantum field theories, see e.g., \cite{Verch:2004vj}. Understanding the nature of entanglement in different quantum states in this context is not only interesting in its own right, but also from the potential connections with holographic dualities. Indeed, the geometrization of the notions of entanglement entropy in the gauge/gravity context for holographic field theories as originally proposed in  \cite{Ryu:2006bv,Ryu:2006ef} (and made geometrically covariant in \cite{Hubeny:2007xt})  makes one wonder if there are further lessons one can learn by understanding the distinct notions of entanglement in the context holographic field theories.

Moreover, the connections between geometry and entanglement as we  now are starting to understand are perhaps much more intimate. The original arguments espoused in 
\cite{Swingle:2009bg} and \cite{VanRaamsdonk:2009ar,VanRaamsdonk:2010pw} suggest a close association between entanglement inherent in a quantum state and the realization of the holographic dual in terms of classical geometry. The relation between entanglement and the emergence of a macroscopic spacetime, is further  bolstered by the arguments of \cite{Maldacena:2013xja} who suggest an intimate connection between EPR like states and Einstein-Rosen bridges, succinctly summarized by the catchphrases ``ER =EPR'' or ``entanglement builds bridges''.\footnote{ See also \cite{Hartman:2013qma} for suggestions relating growth of entanglement  with that of spacetime volume created thus using analogies with tensor networks and \cite{Nozaki:2013vta, Lashkari:2013koa, Bhattacharya:2013bna, Faulkner:2013ica,Swingle:2014uza} for attempts to recover gravitational dynamics from quantum information.}

While these fascinating developments hint at an underlying structure wherein entanglement of quantum states plays an important role in emergence of macroscopic geometry and gravitational physics from the microscopic quantum dynamics, it is fair to ask whether the different notions of entanglement as described above have any useful intuition to impart in explicating the general structure. Does the spacetime geometry care if the entanglement is EPR like, or if it undistillable, or if the quantum entanglement is contaminated by classical correlations? These are, we believe, interesting questions whose answers may potentially shed some light into the geometrization of quantum entanglement.

In this paper, we undertake a modest step in this direction by studying the properties of entanglement negativity, which as previously mentioned is a computable measure of entanglement, in relativistic field theories. We begin in \S\ref{sec:negative} by reviewing the necessary definitions in quantum mechanics and use these to guide our intuition for negativity in simple examples. We first show quite generally that the entanglement negativity of a thermofield double state (the pure entangled state in a tensor product Hilbert space) has a very simple expression in  terms of  the difference of free energies. 

While this result is a corollary of a more general result known already in \cite{Vidal:2002zz} relating the  entanglement negativity of pure states in a bipartite system to a particular Renyi entropy (at index $\frac{1}{2}$) of the reduced density matrix for one of the components, it  casts the general result in simple terms, which in turn allows us to extract some lessons. We argue for instance in \S\ref{sec:cft0} that it allows us to recover the negativity of the vacuum state of a CFTs for a spherical partitioning  of the spatial geometry. In particular, employing the conformal mapping developed in \cite{Casini:2011kv}, we give results for the entanglement negativity for spherical regions for $d$-dimensional CFTs.  In this context it bears mentioning that the results for entanglement negativity have been obtained in $2$-dimensional CFTs by employing the replica trick in \cite{Calabrese:2012ew,Calabrese:2012nk}. These results are of course more powerful and express the computation of the entanglement negativity in terms of twist operator correlation functions for cyclic orbifolds.  In \S\ref{sec:holneg} we make some general comments on extracting the negativity in holographic field theories using the generalized gravitational entropy prescription of \cite{Lewkowycz:2013nqa} and comment on some general lessons one can learn from these analysis. We conclude with a discussion of open questions in \S\ref{sec:discuss}.

\section{Entanglement negativity}
\label{sec:negative}

While our ultimate aim is to explore quantum information theoretic ideas in the holographic realm, we first however need to explain the basic concepts. We therefore begin our discussion with a review of the salient issues relevant for discussing entanglement negativity in quantum mechanics, and postpone generalizations to relativistic quantum field theories to a later stage.

As discussed in \S\ref{sec:intro}, given a density matrix describing a mixed state of some bipartite system it is natural to ask whether there is any way we can reveal if it is separable or entangled. More generally one could  hope to find a criterion to distinguish different kinds of entanglement in general, which could prove useful in various contexts as discussed hitherto. 

A powerful tool in this direction is the so called {\em positive partial transpose} criterion (PPT).
Consider the set-up described in \S\ref{sec:intro} where we have a bipartite system\footnote{ We focus exclusively on bipartite entanglement. Attempts to understand multipartite entanglement in the holographic context can be found in \cite{Balasubramanian:2014hda} (see also \cite{Gharibyan:2013aha}).}   in a tensor product Hilbert space ${\cal H}_L \otimes {\cal H}_R$.   We pick a basis in the space of each subsystem $\ket{{\mathfrak r}_a}$ and $\ket{{\mathfrak l}_\alpha}$ with $a \in \{1, 2,\cdots, \text{dim}({\cal H}_R)\}$ and $\alpha \in \{1, 2,\cdots, \text{dim}({\cal H}_L)\}$, making clear left-right distinction in our notation.
A general density matrix $\rho$ (or indeed any operator ${\cal O}$) in the tensor product ${\cal H}_L \otimes {\cal H}_R$ has matrix elements in our chosen basis
\begin{equation}
\rho_{a\alpha, b \beta}  = \bra{{\mathfrak r}_a \, {\mathfrak l}_\alpha} \!\rho \! \ket{{\mathfrak r}_b\, {\mathfrak l}_\beta} \,.
\label{}
\end{equation}  
On occasion we will need to also talk about the reduced density matrix of one of the components ${\cal H}_{R,L}$. We define then $\rho^{R} = {\rm Tr}_{L} (\rho)$ as the reduced density matrix inherited for the right subsystem from $\rho$ (similarly $\rho^L$). Given such a density matrix, one defines the {\em partial transpose} with respect to the one of the systems, which w.l.o.g. we take to be the left system.\footnote{ With this understanding we denote the partial transpose of $\rho$ by $\ptr{\rho}$, economizing notation by dispensing with indicating that the left subsystem was transposed.} Denoting this partial transposed density matrix as $\ptr{\rho}$ we have its matrix elements in the aforementioned  basis to be
\begin{equation}
\ptr{\rho}_{\!\!\! a\alpha, b \beta}  = \rho_{a\beta, b \alpha} =\bra{{\mathfrak r}_a \, {\mathfrak l}_\beta} \!\rho \! \ket{{\mathfrak r}_b\, {\mathfrak l}_\alpha} \,.
\label{}
\end{equation}  
If $\ptr{\rho}$ has non-negative eigenvalues then $\rho$ is said to have positive partial transpose (PPT). With these definitions one has the following criterion due to Peres \cite{Peres:1996dw}
\begin{center}
{\em  $\rho$ is separable} $\;\; \Longrightarrow\;\;$ $\rho$ {\em is PPT}
\end{center}
The converse is true only for two-qubit (and qubit-trit) systems but not for higher dimensional Hilbert spaces \cite{Horodecki:1996nc}. 

As discussed earlier all distillable states are in direct conflict with local realism, so one could think that only separable states are undistillable. Here the PPT criterion comes strongly into play showing that this intuition is wrong. In fact it was proved in \cite{PhysRevLett.80.5239} that
\begin{center}
{\em $\rho$ is PPT $\;\;\Longrightarrow\;\;$ $\rho$ is undistillable} 
\end{center}
For this reason these states are called {\em bound entangled} in contrast to {\em free entanglement} that can be distilled. In other words if a state is bound entangled it is not possible to extract pure entanglement from it using only LOCC. The authors of \cite{PhysRevLett.80.5239} proposed an interesting analogy with thermodynamics. To prepare a bound entangled state some amount of entanglement is necessary, but the process is irreversible, as after the state is produced it is not possible to distill the initial entanglement back.

It is then reasonable to ask whether a PPT state, which is undistillable, is local in the sense of Werner. Indeed Peres conjectured in \cite{Peres:1998aa} that this is the case, i.e., if a state is PPT it cannot violate {\em any} Bell inequality. The question remained open for fifteen years even if strong evidence has been found in its support (cf., results in \cite{Werner:2000aa,Terhal:2003aa} and references therein). Very recently the conjecture has finally been disproved in \cite{Vertesi:2014aa} where a small violation of a Bell inequality has been found for a particular PPT state. This shows that local Werner states cannot be exactly identified with undistillable states.

As we see above,  while the PPT criterion per se  is not conclusive in identifying local entanglement, it can be used to define a measure of the amount of distillable entanglement contained in a state. This measure, called {\em negativity}, was introduced first in \cite{Vidal:2002zz} and will form the focus of our investigation.

Given  a density matrix $\rho$ one defines  the  {\em negativity} as measure of entanglement based on the amount of violation of the PPT criterion\footnote{ A comment about the notation: the negativities depend not only on the state $\rho$ but also the bipartitioning. We refrain from explicitly indicating the latter to keep the notation clean.} 
\cite{Vidal:2002zz}
\begin{align}
&\text{Negativity:}&  \hspace{-3cm} 
\Neg(\rho) = \frac{||\ptr{\rho}||_1 -1}{2} \,,
\label{defN}
 \\
& \text{Logarithmic Negativity:} &  \hspace{-4cm} 
\Eneg(\rho) = \log ||\ptr{\rho}||_1\,,
\label{defLN}
\end{align}
where $||{\cal O}||_1$ denotes the trace-norm of an operator 
\begin{equation}
||{\cal O}||_1 = \text{Tr} \left(\sqrt{{\cal O}^\dagger \, {\cal O}}\right) .
\label{trnorm}
\end{equation}	
Recall that operationally the trace norm computes the sum of the absolute values of the eignevalues of an operator $ ||{\cal O}||_1 = \sum_i \; |\lambda_{{\cal O},i}| $, i.e., $||{\cal O}||_1 = {\rm Tr} |{\cal O}|$. As a result one is effectively computing a ``signed trace'' with non-trivial weighting for the negative eigenvalues of the partial transposed matrix $\ptr{\rho}$.

For completeness we also recall the notions of entanglement entropy and entanglement Renyi entropies:
\begin{align}
S(\rho) &= -{\rm Tr} \left(\rho \log \rho \right)= \lim_{q\to 1} S^{(q)} (\rho) \,,
\nonumber \\
S^{(q)}(\rho) &= \frac{1}{1-q} \, \log {\rm Tr} \left( \rho^q \right) \,, \qquad \qquad 
q \in {\mathbb Z}_+
\label{eerenyi}
\end{align}

From the definition of the trace norm it then follows that the negativity provides a measure of the number of negative eigenvalues of the density matrix $\ptr{\rho}$. Indeed, passing to a Schmidt basis,  with eigenvalues of $\ptr{\rho}$ being 
$\{\lambda_i^{(+)} , \lambda_j^{(-)},0_k\}$, with the non-zero eigenvalues separated by their parity, we see that 
\begin{equation}
 {\rm Tr} (\ptr{\rho}) 
= \sum_i\, \lambda_i^{(+)} +\sum_j \,\lambda_j^{(-)} \equiv 1 ={\rm Tr} (\rho ) \,.
\label{}
\end{equation}	
Here and in the following we will assume that the density matrix  to be normalized as indicated.  Note that while the eigenvalues of $\ptr{\rho}$ are different from those of $\rho$ generically, the trace is invariant under partial transposition.
On the other hand 
\begin{equation}
\Neg(\rho) 
= \frac{1}{2} \left( \sum_i \, |\lambda_i^{(+)}| + \sum_{j}\, |\lambda_j^{(-)}| -1\right)
= \sum_{j} |\lambda_j^{(-)}| \,,
\label{}
\end{equation}	
is the sum of the absolute values of the negative eigenvalues of $\ptr{\rho}$, explaining the terminology.  At the risk of being pedantic, let us note that the negativity is a property of the original density matrix $\rho$  (the partial transpose $\ptr{\rho}$  is just a computational aid).

Properties of negativities have been discussed in the literature on quantum information, cf., \cite{Vidal:2002zz,Audenaert:2003aa,Plenio:2005aa,He:2014aa,Ou:2007aa}. By construction both the negativity and the logarithmic negativity fail to detect bound entangled states and for this reason they do not quantify the total amount of entanglement inherent in a mixed state of the system. Furthermore it is important to note that even in case of pure states these quantities do not in general agree with entanglement entropy. Specifically, the logarithmic negativity gives in general a larger measure of entanglement, as we will see explicitly below, while the negativity reduces to one half of the {\em entanglement robustness}\footnote{The robustness of entanglement can be understood as a measure of the amount of noise required to disrupt the entanglement of the system. See \cite{Vidal:1999aa} for more details.}. Agreement with entanglement entropy on pure states is a property commonly required in the construction of an axiomatic entanglement measure\footnote{For more details on the axioms that have to be satisfied by entanglement measures see \cite{Plenio:2007aa} .}, but the case of negativities is different. This fact distinguishes negativities from most entanglement measures, such as entanglement of formation and distillation, which instead reduce to entanglement entropy for pure states. Nevertheless negativities can be used to quantify entanglement provided that they do not increase under any LOCC, i.e., they are {\em entanglement monotones}. This is indeed the case, as proved in \citep{Plenio:2005aa}\footnote{It was actually shown that negativities are entanglement monotones under a larger class of operations called PPT-operations. This is the class of all operations that map the set of PPT states to itself. For further properties of negativities relatively to PPT-operations see \citep{Audenaert:2003aa}.}.

The previous properties are shared by both negativities, but each of them has peculiar properties of interest by itself. The logarithmic negativity for example has been shown to give an upper bound to the distillable entanglement of $\rho$ and to satisfy an additivity relation. For a separable state of a bipartite system of two parties $A$ and $B$ one indeed has
\begin{equation}
\Eneg(\rho_A\otimes\rho_B)=\Eneg(\rho_A)+\Eneg(\rho_B)\,.
\label{negfactor}
\end{equation}
On the other hand the negativity can be related to the maximal fidelity that can be achieved in a teleportation protocol that uses $\rho$ as a resource. 

It is interesting to note that the negativity satisfies an interesting disentangling theorem. Consider a tripartite system $ABC$ in a pure state $\ket{\Psi_{ABC}}$ and denote the negativity between $A$ and $BC$ as $\Neg_{A|BC}$ and the negativity between $A$ and $B$ as $\Neg_{A|B}$. It was recently proved in \citep{He:2014aa} that if and only if 
$\Neg_{A|BC}=\Neg_{A|B}$  then there exists a  partitioning of $B$ into $B_1$ and $B_2$ such that the state of the whole system factorizes
\begin{equation}
\ket{\Psi_{ABC}}=\ket{\Psi_{AB_1}}\otimes\ket{\Psi_{B_2C}} 
\label{trineg1}
\end{equation}
It is an immediate consequence that under the hypothesis of the theorem the negativity between $A$ and $C$ (denoted as $\Neg_{A|C}$) is zero, equivalently the reduced density matrix obtained from $\ket{\Psi_{ABC}}$ by tracing out $B$ factorizes: $\rho_{AC}=\rho_A\otimes\rho_C$. Furthermore, in this particular case, one has the saturation of a monogamy inequality for the square of the negativity previously proved by \cite{Ou:2007aa} for systems of three qubits
\begin{equation}
\Neg^2_{A|BC}\geq\Neg^2_{A|B}+\Neg^2_{A|C}
\label{trineg2}
\end{equation}
The authors of \citep{He:2014aa} conjectured that this inequality should be true in general giving numerical results in its support. Finally it interesting to mention that contrary to what one could expect, the previous inequality does not hold for the negativity itself.

To build some intuition for the negativity, we would like to understand its properties in simple situations. It should be no surprise to the reader that negativity can be non-vanishing even in pure states. After all the simple Bell state for a 2-qubit system we can have an EPR state $\frac{1}{\sqrt{2}} \left(\ket{\,\uparrow \uparrow} + \ket{\,\downarrow \downarrow} \right)$ which is pure, but entangled. It is easy to verify that for this state the negativity is $\frac{1}{2}$. Perhaps more usefully, the 
logarithmic negativity is $\log 2$ which is also the von Neumann entropy for the reduced density matrix for one of the qubits.  It is easy to see that this result is not restricted to two-qubit systems. One has the following general result:

\noindent 
{\em Logarithmic entaglement negativity of a maximally entangled state $\psi_{\text{max}}$ in a bipartite system equals the Entanglement entropy of  the reduced density matrix $\rho_{\text{max}}^{L,R}$ for one of the subsystems.} 
\begin{equation}
\Eneg(\psi_{\text max}) = S(\rho_{\text max}^{R,L}) \,.
\label{negmax}
\end{equation}	
While this statement illustrates the basic feature of this particular measure of entanglement it is useful to look at a simple generalization that will allow us to build some intuition for the negativity.\footnote{ We find it convenient to notationally distinguish pure and mixed states and therefore denote a pure state density matrix as $\psi = \ket{\Psi} \bra {\Psi}$.}

\subsection{Negativity in thermofield state}
\label{sec:thermal}

 Let us consider the thermofield double state $\ket{\Psi}_\beta$ in with ${\cal H}_{L,R}$ being two copies of the same physical system. Working in an energy eigenbasis with spectrum $\{E_i\}$ we have\footnote{ We have for simplicity assumed that we are dealing with a finite system where $\text{dim}({\cal H}_L)  = \text{dim}( {\cal H}_R) = N$.}
\begin{equation}
\ket{\Psi}_\beta = \frac{1}{\sqrt{Z(\beta)}}\, \sum_{a=1}^{N} \, e^{-\frac{\beta}{2}\, E_a} \ket{{\mathfrak r}_a \, {\mathfrak l}_a} 
\label{}
\end{equation}	
The state in the tensor product is of course pure, but entangled. We want to take a measure of this entanglement, using the logarithmic negativity $\Eneg(\psi_\beta)$ with 
\begin{equation}
\psi_\beta =\; \ket{\Psi}_\beta \, \bra{\Psi}_\beta \; = 
\frac{1}{Z(\beta)} \sum_{a,b=1}^N \; e^{-\frac{\beta}{2} (E_a + E_b)} \, \ket{{\mathfrak r}_a \, {\mathfrak l}_a} \; \bra{{\mathfrak r}_b \, {\mathfrak l}_b} 
\label{}
\end{equation}	
It is then trivial to see that 
\begin{equation}
\ptr{\psi}_\beta = \frac{1}{Z(\beta)} \sum_{a,b=1}^N \; e^{-\frac{\beta}{2} (E_a + E_b)} \, \ket{{\mathfrak r}_a \, {\mathfrak l}_b} \; \bra{{\mathfrak r}_b \, {\mathfrak l}_a} 
\label{}
\end{equation}	
and 
\begin{equation}
\ptr{\psi}_{\beta}^{\dagger} \, \ptr{\psi}_\beta  = \frac{1}{Z(\beta)^2} \; \sum_{a,b=1}^N \; e^{-\beta (E_a + E_b)} \, \ket{{\mathfrak r}_a \, {\mathfrak l}_a} \; \bra{{\mathfrak r}_b \, {\mathfrak l}_b}
\label{}
\end{equation}	
whence it follows that 
\begin{equation}
\Eneg(\psi_\beta) = \log \frac{Z(\frac{\beta}{2})^2}{Z(\beta)} = \beta \left( F(\beta) - F(\beta/2) \right)
\label{negthermal}
\end{equation}	
with the final result written in terms of the free energy $F(\beta) = - \frac{1}{\beta} \, \log Z(\beta)$. 

The logarithmic negativity of the thermofield state $\psi_\beta$ is proportional  to the difference of free energies of the system at temperature $T$ and $2\,T$ respectively.\footnote{ The simplicity of the final result in terms of the free energy difference is the reason for preferring the logarithmic negativity over the negativity itself. We henceforth will focus on $\Eneg$ and refer to it as the negativity in the rest of our discussion for convenience.}  This is main observation which we will exploit in the sequel to obtain some insight into the nature of entanglement in quantum field theories. On the other hand the reduced density matrix $\rho^{R,L}_\beta$ for the right or left systems has a von Neumann entropy $S(\rho_{\cal A})$ which is obtained directly from $Z(\beta)$ itself. In the limit $\beta \to 0$ we recover the previous assertion \eqref{negmax} for maximally entangled states.

\subsection{Renyi  Negativities}
\label{sec:renneg}

For the thermofield state there is a simple relation between the negativity of the total density matrix and the reduced matrix of one component. This in fact generalizes to pure states of the bipartite system quite simply. To get further intuition for the negativity, it is worthwhile to follow the line of thought that led to the replica analysis for entanglement entropy. Just as we consider the moments of the density matrix in order to compute its von Neumann entropy, we now examine the moments of the partial transpose $\ptr{\rho}$.

Consider following \cite{Calabrese:2012ew} the notion of Renyi negativity for a density matrix $\rho$:
\begin{align}
\exp\left(\Eneg^{(q)}(\rho) \right)= {\rm Tr} \,(\ptr{\rho})^{\,q}  = 
\begin{cases}
& \sum_i \, \left(\lambda_i^{(+)}\right)^{q_e} + \sum_j \left(\lambda_j^{(-)}\right)^{q_e} \,, 
\quad q_e \in 2\,{\mathbb Z}_+ \cr
&  \sum_i \, \left(\lambda_i^{(+)}\right)^{q_o} - \sum_j \left(\lambda_j^{(-)}\right)^{q_o} \,,
\quad q_o \in 2\,{\mathbb Z}_+ +1
\end{cases}
\label{renyiE}
\end{align}
As is clear from the above definition the parity of the integer $q$ plays a crucial role. Should we wish to employ the replica construction and  recover the logarithmic negativity from these Renyi entropies then we will need to exclusively use the even sequence. The logarithmic negativity is obtained by an analytic continuation of even Renyi negativities to $q_e \to 1$, i.e.,
\begin{equation}
\Eneg(\rho)  = \lim_{q_e \to 1} \,  \Eneg^{(q_e)} \,, \qquad q_e \in 2{\mathbb Z}_+
\label{}
\end{equation}	

Using the definition \eqref{renyiE} we can immediately generalize our considerations for the thermal state to any pure state $\psi = \ket{\Psi} \bra{\Psi}$ of a bipartite system. We have 
\cite{Calabrese:2012nk} 
\begin{align}
\Eneg^{(q_e)}(\psi) &= 2\,(1-\frac{q_e}{2})\; S^{(q_e/2)}(\rho^{R,L}) \,,
\nonumber \\
\Eneg^{(q_o)}(\psi) &=  (1-q_0) \;S^{(q_o)}(\rho^{R,L}) \,.
\label{negrenyi}
\end{align}	
In particular note that 
\begin{equation}
\Eneg(\psi) =  S^{(1/2)}(\rho^{R,L}) \,,
\label{renyi12}
\end{equation}	
as the generalization of our previous assertions \eqref{negmax} and \eqref{negthermal}.
We note that the Renyi negativities have been used to extract the negativities in two dimensional conformal field theories (CFTs) in \cite{Calabrese:2012ew,Calabrese:2012nk}. The technical tool involved is to appropriately map the computation as in the case of entanglement entropy to that of computing twist operator correlation functions. We will have occasion to comment on some of their results in due course.

\section{Negativity of a CFT vacuum}
\label{sec:cft0}

Having defined the basic quantity of interest let us now turn to its computation in relativistic field theories. To our knowledge the only study of negativity in such as context are the aforementioned works \cite{Calabrese:2012ew,Calabrese:2012nk} who examine its behaviour in $2d$ CFTs. Our interest is in understanding properties of negativity more generally. In what follows we explain how one can exploit \eqref{negthermal} to find explicit results for a certain choice of bipartitioning of the vacuum state of a CFT. Subsequently we describe how to tackle the problem more generally.

Consider a relativistic QFT in $d$-dimensions on some background geometry ${\cal B}$. As remarked earlier in \S\ref{sec:intro} we want to ask how to quantify the entanglement of the vacuum state in this theory. For the present we are going to use the concept of logarithmic  negativity introduced in \S\ref{sec:negative} to serve as the measure of interest. 

A natural way to proceed  is to consider a spatial Cauchy slice $\Sigma$  and consider some region ${\cal A}\subset \Sigma$. One can ask how degrees of freedom in ${\cal A}$ are entangled with those in ${\cal A}^c = \Sigma\backslash {\cal A}$. By now we have a good idea about the entanglement entropy associated with the reduced density matrix $\rho_{\cal A} = {\rm Tr}_{{\cal A}^c} (\ket{0}\bra{0})$ either by direct field theory computation in $d=2$ using the replica trick or using holography in all $d$.

To be be specific let us examine two situations which are particularly simple, where the field theory calculation boils down effectively to a spectral computation. Consider a conformally invariant field theory which we will place on one of two background geometries for the present:
\begin{enumerate}
\item[(i).] ${\cal B}_d = {\mathbb R}^{d-1,1}$ (Mink): $\ket{0_p\,}$ is the Minkowski or Poincar\'e vacuum and  ${\cal A}$ is a ball shaped region centered w.l.o.g. at the origin
\begin{align}
{\cal A} \subset {\mathbb R}^{d-1}: \quad r \leq R \,, \qquad ds^2_{\cal B} = -dt^2 + dr^2 + r^2 \, d\Omega_{d-2}^2\,.
\label{}
\end{align}
\item[(ii).]  ${\cal B}_d = {\bf S}^{d-1} \times {\mathbb R}$ (ESU). $\ket{0_g\,}$ is the global or vacuum and  ${\cal A}$ is a polar-cap region about the north pole of ${\bf S}^{d-1}$
\begin{align}
{\cal A} \subset {\bf S}^{d-1}: \quad \theta \leq \theta_{\cal A} 
\,, \qquad 
ds^2_{\cal B} = -dt^2 + R^2 \left(d\theta^2 + \sin^2\theta\, d\Omega_{d-2}^2\right).
\label{}
\end{align}
\end{enumerate}
The reasons for using $R$ to denote the size of the ball as well as the curvature radius of the sphere in the two distinct cases will become clear momentarily.  For these two cases we will exploit a well known fact about the reduced density matrix 
$\rho_{\cal A}$ to make some inferences about the negativity.

Let us begin by recalling some salient features elucidated in \cite{Casini:2011kv}. For our two regions the domain of dependence $\domd \subset {\cal B}$ is conformally equivalent to the hyperbolic cylinder ${\mathbb H}_d = H_{d-1} \times {\mathbb R}$, with the curvature radius of the hyperbolic space $H_{d-1}$ being $R$.  Since the entanglement structure is a property of an entire causal domain, not just a spatial region, we can as well think of $\Eneg(\rho_{\cal A})$  as a function defined on $\domd$.\footnote{ In the language of \cite{Headrick:2014sf} we should think of the negativity also as a wedge observable. Thus it is also subject to the constraints of causality as described therein for entanglement entropy.} 

With this understanding the conformal mapping of \cite{Casini:2011kv} implies that  the reduced density matrix $\rho_{\cal A}$ is unitarily equivalent to the thermal density matrix for the CFT on the hyperbolic cylinder\footnote{ We refer the reader to \cite{Casini:2011kv} for explicit expressions of the unitary map.} ${\mathbb H}_d$
\begin{equation}
\rho_{\cal A} = {\cal U} \, \rho_\beta \, {\cal U}^\dagger \,,  \qquad \beta = 2\,\pi\, R \,.
\label{thermalH}
\end{equation}	
We note that the temperature is set by $R$ and in particular it is independent of $\theta_{\cal A}$ for the theory on ESU.  This is intuitive on dimensional grounds, though we should note that the angular dependence is implicit in $\rho_{\cal A}$. For e.g., in computing entanglement entropy a $\theta_{\cal A}$ dependence will arise by relating the UV cut-off on ESU with the IR cut-off for the CFT on the hyperbolic cylinder. It is perhaps more instructive to note that the modular Hamiltonian defined via $ \rho_{\cal A}= e^{-H_{\cal A}} $ has an explicit dependence on the angular extent of the polar-cap (see e.g., \cite{Gentle:2013fma}).

We interpret this result as follows. The vacuum state of the tensor product  
${\cal H}_{\cal A} \otimes {\cal H}_{{\cal A}^c}$ for the aforementioned choice of regions is expressible in terms of the thermal state on the hyperbolic cylinder. Schematically, we can write
\begin{equation}
\psi_0 \big|_{\text{Mink, ESU}}  = \psi_\beta \big|_{{\mathbb H}} \,,
\label{}
\end{equation}	
From this observation using \eqref{negthermal} we infer that  (for either $\ket{0_p}$ or $\ket{0_g}$)
\begin{equation}
\Eneg(\psi_0) = 2\,\pi\,R \left( F_{\mathbb H}(2\pi R) - F_{\mathbb H}(\pi R) \right),
\label{}
\end{equation}	
where $F_{\mathbb H}$ is the free energy of the CFT on the hyperbolic cylinder. 

So the problem of computing negativity in the vacuum state of a CFT can thus be mapped to computing the spectrum on the hyperbolic space. As long as we have this spectral data we can then immediately infer the negativity of the vacuum. It will turn out that the negativity has an inherent UV divergence and necessitates a UV regulator for its computation. 

To ascertain the divergence structure we note that a UV regulator on ${\cal B}_d$ maps to an IR regulator on ${\mathbb H}_d$ by virtue of the conformal mapping. We have from the analysis of  \cite{Casini:2011kv} the relations 
\begin{equation}
L_{\mathbb H} = \log\left(\frac{2R}{\epsilon_\text{Mink}} \right) 
\,, \qquad 
L_{\mathbb H}  = \log\left(\frac{2\,R}{\epsilon_\text{ESU}} \,\sin\theta_{\cal A}\right),
\label{Hcutoff}
\end{equation}  
in the two cases of interest. Here $L_{\mathbb H}$ is the IR regulator of the length scale in the hyperbolic cylinder and $\epsilon_{\cal B}$ is the UV cut-off in the background indicated.
This mapping between the cut-offs can be used to express the volume of the hyperbolic cylinder in terms of field theory data on ${\cal B}_d$. Denoting by 
 $ \text{Vol}(H_{d-1})$ the spatial volume of a unit radius of curvature hyperbolic space,using the explicit expression mapping the cut-offs  \eqref{Hcutoff}, one obtains the desired expression for ${\cal B}_d= \text{Mink}_d$, 
\begin{align}
 \text{Vol}(H_{d-1})&= \omega_{d-2} \, \int_1^{\frac{R}{\epsilon}} \, dx \, (x^2 -1)^\frac{d-3}{2}  
\,, \qquad \quad \omega_{d-2} = \frac{2\,\pi^\frac{d-1}{2}}{\Gamma\left(\frac{d-1}{2} \right)} 
\nonumber \\
&\simeq 
\frac{\omega_{d-2}}{d-2}\, \left[  \left(\frac{R}{\epsilon}\right)^{d-2} - 
\frac{(d-2)(d-3)}{2\,(d-4)}\, \left(\frac{R}{\epsilon}\right)^{d-4} + \cdots + V_\text{univ}\right]
\label{Hvol}
\end{align}
where 
\begin{align}
V_\text{univ} = \dfrac{\sqrt{\pi}}{2}\; \dfrac{\Gamma(\frac{d-1}{2})}{\Gamma(\frac{d}{2})}  
\begin{cases}
&  (-1)^{\frac{d}{2}-1}\, \dfrac{2}{\pi} \, 
\log\left(\dfrac{2\,R}{\epsilon} \right)  \,, \qquad d\in 2{\mathbb Z}_+\cr 
& 
 (-1)^{\frac{d-1}{2}}    \,, 
 \qquad \hspace{2.5cm} d \in 2{\mathbb Z}_+ +1
\end{cases}
\label{}
\end{align}  
Similar expressions can be derived for ${\cal B}_d = \text{ESU}_d$; all we would need to do is replace the upper limit of the integral in \eqref{Hvol} by the appropriate cut-off expression  given in \eqref{Hcutoff}. Armed with this information we now present some expressions for the negativity using various results already present in the literature.

\paragraph{CFTs in 2 dimensions:}  In $d=2$ we have a simplifying feature that $H_1$ is flat. Indeed using the result $F(T) = - \frac{\pi}{12}\, (c_L + c_R)\, T^2 \, L$ for a thermal CFT at temperature $T$ in spatial volume $L$ we find
\begin{align}
\Eneg(\psi_0) &=  \frac{c}{2}\, \log X  \,, \qquad 
X =
\begin{cases}
&\frac{2\,R}{\epsilon}   \,,\qquad {\cal B} = \text{Mink} \\
& \frac{2\, R}{\epsilon}\,\sin \theta_{\cal A}  \,, \qquad {\cal B}= \text{ESU} 
\end{cases}
\label{}
\end{align} 
One may alternatively have derived this answer by using \eqref{renyi12} and the familiar result $S^{(q)} = \frac{c}{6}\, \left(1+\frac{1}{q}\right)\, \log X$ for CFT$_2$.

\paragraph{Free CFTs in various dimensions:} The second example where we can explicitly compute the negativity is to use the results for the free energy $F_{\mathbb H}$ of free fields 
in various dimensions. Results for free scalars in all dimensions were derived initially in \cite{Casini:2010kt} and analogous results for various theories  in $d=3$ were obtained in  \cite{Klebanov:2011uf}. From here we can immediately read off the answer for the Renyi entropy at $q=\frac{1}{2}$ and thence the negativity using \eqref{renyi12}. 

For a free field of mass $m$ in ${\mathbb R}^{2,1}$ one has the free energy on 
${\mathbb H}$ at temperature $\beta$ explicitly in closed form \cite{Klebanov:2011uf} in terms of  the function
\begin{align}
{\cal I}_{\eta,q}(m) =  \int_0^\infty\, d\lambda \, \lambda \, \tanh^\eta(\pi\,{\lambda}) \, 
\log\left(1 -\eta\, e^{-2\pi\,q\sqrt{\lambda^2+m^2}} \right) .
\label{}
\end{align}	
Here $\eta = \pm 1$ encode the statistics ($\eta = +1$ for bosons and $\eta  = -1$ for fermions respectively).  One then  finds that the negativity for free massless fields are given as 
\begin{align}
\Eneg(\psi_p^\eta) 
&= \frac{\text{Vol}(H_{2})}{\pi} \;  
\left( {\cal I}_{\eta,1}(0) -2 \,{\cal I}_{\eta,\frac{1}{2}}(0) \right)
\nonumber \\
&= 
\frac{\text{Vol}(H_{2})}{\pi} 
\int_0^\infty\, d\lambda \, \lambda\, \tanh^\eta(\pi\,{\lambda}) \, 
\log\left(
\frac{1- \eta\, e^{-2\pi\,\lambda} }
{\left(1-\eta\, e^{-\pi\,\lambda} \right)^2}\right).
\label{negfree}
\end{align}	
Note that the integral is convergent and all the divergences are encoded in the pre-factor 
$\text{Vol}(H_2)$, which we have already expressed in terms of the relevant variables in  Eq.~\eqref{Hvol}. The expression for $\Eneg(\psi_g)$ would be similar with an appropriate replacement of the volume of the hyperbolic space.

Let us also record the  expression for the entanglement entropy for the reduced density matrix $\rho_{\cal A}$ for comparison. One has from  \cite{Klebanov:2011uf}
\begin{align}
S(\rho^\eta_{\cal A})  &= 
 \frac{\text{Vol}(H_{2})}{2\pi} 
\left[{\cal I}_{\eta,1}(0)-  \frac{(7-\eta)\,\zeta(3)}{8\,\pi^2}\right]\,.
\label{entfree}
\end{align}

We  see from \eqref{negfree} and \eqref{entfree} that the divergent terms in the negativity are (structurally) the same as in the entanglement entropy; the numerical coefficient however is rather different. Let us define the ratio 
\begin{equation}
{\cal X}_d = \bigg| \frac{\tt{C}_\text{univ} \left[\Eneg(\psi_p)\right]}{
{\tt C}_\text{univ} \left[S(\rho_{\cal A})\right]} \bigg|
\label{Xdef}
\end{equation}	
where ${\tt C}_\text{univ}[x]$ denotes the coefficient of the universal term $V_\text{univ}$ occurring in the expression $x$. We claim that this quantity gives a precise measure of the negativity  for $\ket{\!0}$  in terms of the entanglement entropy of the reduced density matrix $\rho_{\cal A}$. 

For a free massless scalar in $d=3$ we find ${\cal X}_3^\text{free}\approx 2.716$, while for a massless fermion ${\cal X}_3^\text{free}\approx 1.888$. We note that ${\cal X}_3(m)$ defined formally for massive fields is a monotonically increasing function of $m$. We will return to this ratio below once we also obtain analogous results from holography for strongly coupled CFTs.

Results for Renyi entropies for spherical entangling regions are also known for free $SU(N)$ 
${\cal N}=4$ Super-Yang Mills theory in $d=4$ \cite{Fursaev:2012mp}. From these results we find
\begin{align}
\Eneg(\psi_p) &\simeq N^2\, \left[\frac{R^2}{\epsilon^2} - 
\frac{41}{24}\, \log\left(\frac{R}{\epsilon}\right) \right] 
\nonumber \\
S(\rho_{\cal A}) &\simeq  N^2\, \left[\frac{1}{2}\, \frac{R^2}{\epsilon^2} -  \log\left(\frac{R}{\epsilon}\right) \right] 
\label{N4results}
\end{align}
This is a peculiar example where the structure of divergent terms in the negativity for the ground state differs from that in the entanglement entropy of the reduced density matrix induced in the spherical region ${\cal A}$.\footnote{ We find this rather peculiar in light of the conformal mapping described above. Given the free scalar/fermion and holographic results one might have been tempted to consider the ratio of the negativity to the entanglement entropy en masse, without isolating the universal part (assuming both computations be regulated in a similar fashion). We thank Horacio Casini and Tadashi Takayanagi for useful discussions on this point.} From the expressions above we find that ${\cal X}_4 = \frac{41}{24}\simeq 1.708$ for free  ${\cal N}=4$ SYM. 

\paragraph{Holographic CFTs in diverse dimensions:} Our final example is the class of holographic field theories in various dimensions. While the computation of the spectrum of an interacting CFT on ${\mathbb H}$ is in general unfeasible, holography provides us with a simple answer when the CFTs in question have (a) large central charge and (b) a sufficient gap in the spectrum. The reason is that the computation of the free energy at temperature 
$\beta$ amounts to finding an asymptotically locally AdS$_{d+1}$ geometry whose boundary is ${\mathbb H}_d$, with the Euclidean time direction having a period $\beta$. The relevant geometry is well known, it is the so called hyperbolic black hole in AdS$_{d+1}$ 
\cite{Emparan:1999gf}. The bulk metric is given as
\begin{equation}
ds^2 = -\frac{\ell_\text{AdS}^2}{R^2} \, f(r)\, dt^2 + \frac{dr^2}{f(r)} + r^2 \, d\Sigma_{H_{d-1}}^2\,, \qquad f(r) = \frac{r^2}{\ell_\text{AdS}^2} - 
\left(\frac{r_+}{r}\right)^{d-2} \left(\frac{r_+^2}{\ell_\text{AdS}^2}-1\right)- 1
\label{}
\end{equation}  
whose conformal boundary is indeed ${\mathbb H}$ with the desired spatial curvature $R$.
$r_+$ is the location of the horizon and we have explicitly kept the AdS length scale $\ell_\text{AdS}$. We note that the combination of this length scale and the $(d+1)-$dimensional Newton's constant $G^{(d+1)}_N$ gives the effective central charge $c_\text{eff}$ of the dual CFT: $c_\text{eff} = \frac{\ell_\text{AdS}^{d-1}}{16\,\pi\, G_N^{(d+1)}}$.

This geometry has in fact been used before to compute the Renyi entropies for holographic field theories in \cite{Hung:2011nu} and we can in fact use their results to infer the behaviour of the negativity directly. We first note that the black hole thermodynamic data are given in terms of the geometric parameters as
\begin{equation}
T = \frac{d\,r_+^2 - (d-2)\,\ell_\text{AdS}^2}{4\,\pi\, R\, \ell_\text{AdS}\,r_+} 
\,, \qquad
S = \frac{1}{4\,G_N^{d+1}}\, r_+^{d-1}\, \text{Vol}(H_{d-1})\,.
\label{}
\end{equation}  
Given that we know the free energy and the entropy, we can invoke standard thermodynamic relation $S(T) = - \frac{\partial F}{\partial T}$ to obtain the final result \cite{Hung:2011nu}
\begin{align}
\Eneg(\psi_p) &= \pi\, c_\text{eff} \, \text{Vol}(H_{d-1}) \,{\cal X}_d^\text{hol} = S(\rho_{\cal A}) \, {\cal X}^\text{hol}_d\,,
\label{}
\end{align}  
where the dimension dependent coefficient ${\cal X}_d^\text{hol}$ for holographic CFTs is a simple function of the spacetime dimension 
\begin{align}
{\cal X}_d^\text{hol} = \left(\frac{1}{2}\, x_d^{d-2} \, (1+x_d^2)  -1  \right) 
 \,,\qquad x_d = \frac{2}{d} \left(1+ \sqrt{1-\frac{d}{2}+\frac{d^2}{4} }\right) .
\label{}
\end{align}
This function interpolates rather mildly between ${\cal X}_2^\text{hol} =\frac{3}{2}$ and 
$\lim_{d\to \infty} {\cal X}_d^\text{hol} =  (e-1)\approx 1.718$, hinting that up to an overall multiplicative renormalization much of the information is already contained in the entanglement entropy.

It is also curious to note that in $d=3$ one can compare the free field answers to the strong coupling results obtained above.\footnote{ Since we are considering ratios of the negativity to the entanglement,  the precise normalization of central charge $c_\text{eff}$ is immaterial, unlike the case when we compare the entanglement entropy at weak and strong coupling.} For a free scalar field we find ${\cal X}_3^\text{hol}  \approx 0.601 \, {\cal X}^\text{free}_3$, while for a free Dirac field the proportionality is larger
${\cal X}_3^\text{hol}  \approx 0.864 \, {\cal X}^\text{free}_3$.

It would be interesting to understand this ratio which suggests a decrease in (distillable?) entanglement in the strong coupling regime from first principles. The ratio of our measure at weak and strong couplings $\frac{{\cal X}^\text{hol}}{{\cal X}^\text{free}}$ can decrease either by the total entanglement being reduced at strong coupling or more simply by just the negativity decreasing. In the latter case one would only find a decrease in the amount of distillable entanglement at strong  coupling. Ascertaining which of these scenarios is realized might provide new clues in the relation between geometry and entanglement. 

 A similar comparison for ${\cal N}=4$ SYM gives a much more intriguing result ${\cal X}_4^\text{hol}  \approx 0.98 \, {\cal X}^\text{free}_4$, where we switched to using the ratio of the coefficient of the universal logarithmic terms 
\eqref{Xdef} owing to the non-trivial behaviour of the free theory answer \eqref{N4results}. 
 In this case it is rather curious that the free field result undergoes a very mild reduction as we crank up the coupling. Similar comparisons for the Renyi entropies of ${\cal N}= 4$ SYM  at different $q$ are described in some detail in \cite{Galante:2013wta}.

\medskip
\noindent
{\em Note added in v2:} Using the results of \cite{Hung:2011nu} one can compute ${\cal X}_d^\text{hol}$ in quasi-topological theories of gravity. These have been used in the literature to model field theories with unequal central charges (e.g., $a \neq c$ in $d=4$).\footnote{{\em Caveat lector}: While the quasi-topological theories provide a dial to decouple the central charges in large $c_\text{eff}$ theories, we believe they are unphysical, and that there is no unitary field theory whose dual is given precisely by such a gravitational Lagrangian. Rather they should be treated as in any effective field theory as the leading terms in a pertubation expansion of higher derivative operators. We leave it to the reader to decide on the import of the present text which is included to satisfy the curiousity of an anonymous referee.}  The ratio for Gauss-Bonet theory in $d=4$ can be expressed in terms of the $a$ and $c$ central charges as 
\begin{align}
{\cal X}^{hol}_{d=4}(\tilde c) \bigg|_\text{GB} = \frac{x_4^2-1}{8} \, 
\left( (5\, {\tilde c} - 1)\,x_4^2 - (13 \,{\tilde c}-5) 
+ 16\, {\tilde c}\; \frac{2\,{\tilde c} \, x_4^2 - {\tilde c} +1}{ (3\,{\tilde c}-1)\, x_4^2- {\tilde c}+1} \right)
\end{align}
where ${\tilde c} = \frac{c}{a}$ and $x_4$ now solves a cubic equation:
\begin{equation}
x_4^3 - \frac{3\, {\tilde c}-1}{ 5\, {\tilde c} -1}\, (2\, x_4^2 + x_4) + 2\, \frac{{\tilde c} -1}{5\, {\tilde c}-1}=0
\label{}
\end{equation}	
From field theory unitarity considerations bound ${\tilde c} \in [\frac{2}{3},2]$. It is easy to numerically check that 
${\cal X}^{hol}_{d=4}(\tilde c) \bigg|_\text{GB} $ monotonically increases and ranges between $1.397$ and $2.53$ at the ends of the allowed interval.

\section{Holographic negativity: general expectations}
\label{sec:holneg}

Having understood the basic features of entanglement negativity in the vacuum state of a CFT for bipartitioning  by spherical regions, we now turn to more general situations. Most of the discussion below will be of a qualitative nature, devoted to explaining some of the general features.

\subsection{Arbitrary bipartitions of pure states}
\label{sec:genpure}

Let us start with pure states $\ket{\Psi}$. Once again we can focus on bipartitioning a Cauchy slice of the background geometry for the field theory as $\Sigma = {\cal A}\cup {\cal A}^c$. We can then relate the negativity $\Eneg(\psi)$ to the Renyi entropy $S^{(1/2)}(\rho_{\cal A})$ (for the bipartition ${\cal H}_{\cal A}\cup {\cal H}_{{\cal A}^c}$).  Hence as long as we are in a position to compute the Renyi entropies for non-integral values, we would be able to ascertain the negativity. 

To obtain the Renyi entropy at index half, we follow the the holographic computation of \cite{Lewkowycz:2013nqa}.\footnote{ At this stage we have to restrict states $\ket{\Psi}$ to have a moment of time reflection symmetry and  at this preferred instant of time. A general prescription for computing holographic Renyi entropies (even for integer $q$) in time-dependent states is not available at present.} For an arbitrary region ${\cal A}$ we therefore consider replicating the background geometry ${\cal B}$ to ${\cal B}_q$ on which we place our field theory. ${\cal B}_q$ would as usual be characterized by having branch points inside ${\cal A}$ (and its images under the replica construction). Having obtained the answers for integral $q$ which involves finding bulk saddle points with boundary ${\cal B}_q$ we then analytically continue to $q=\frac{1}{2}$. A-priori it is not clear that this last step can be carried out for all choices of ${\cal A}$. 

One can infer the following about the negativity in pure states of a CFT from the basic definition even in the absence of an explicit computation:
\begin{itemize}
\item The negativity in a pure state is divergent with the leading divergent term scaling like the area of the entangling surface $\partial{\cal A}$.
\item The structure of the sub-leading divergent terms is identical to that encountered in the computation of the entanglement entropy for the reduced density matrix $\rho_{\cal A}$ in holographic field theories. This follows from the fact that the divergent terms encountered in the evaluation of the on-shell action in gravity during the computation of the Renyi entropies is independent of $q$. 
\item Perhaps more importantly the value of the negativity $\Eneg(\psi)$ is in general  larger than the entanglement entropy $S(\rho_{\cal A})$. The difference we conjecture should be in a geometric factor.  To wit, the ratio ${\cal X}_{\cal A}$ defined analogously to \eqref{Xdef} should depend just on the geometry of the entangling surface $\partial{\cal A}$.
\end{itemize}

\subsection{Mixed state negativity}
\label{sec:mixed}

 In principle in the holographic discussion we do not need to restrict attention to pure states.  In fact, given that the negativity is naturally intended to test mixed states, one ought to be considering density matrices $\rho$ and attempt to compute their negativity. This as far as we know has been only achieved in $d=2$ CFTs  in \cite{Calabrese:2012nk}. While we have no concrete computation to report in this context, it is worth recording various cases of interest for future exploration.

The general situation which one can consider can be motivated in the following manner. Given a state in some quantum field theory, we focus on some region  ${\cal A}$ lying on a particular Cauchy slice. By integrating out the degrees of freedom in ${\cal A}^c = \Sigma\backslash {\cal A}$ we obtain the reduced density matrix $\rho_{\cal A}$ as usual. Now we further bipartition ${\cal A}$ itself, i.e., divide  ${\cal A}  = {\cal A}_L \cup {\cal A}_R$. With this decomposition at hand we define the negativity  $\Eneg(\rho_{\cal A})$ as before by partial transposing the part of the density matrix associated with ${\cal A}_L$.
As concrete examples consider:
\begin{enumerate}
\item[(a).] Take ${\cal A}$ to be the spherical region of size $R$ in ${\mathbb R}^{d-1}$ considered in \S\ref{sec:cft0} in our previous construction and pick any two mutually adjoining regions for ${\cal A}_L$ and ${\cal A}_R$ respectively. 
 \item[(b).] ${\cal A}$ itself could be the composed of two disconnected regions which we can associate with the bipartitioning of interest.  
\item[(c).] ${\cal A}\subset \Sigma_R$ in the thermofield double state $\ket{\Psi}_\beta \in 
{\cal H}_L \otimes {\cal H}_R$. One can attempt to quantify the negativity of $\rho_\beta^R$ for the bipartition defined by $\Sigma_R = {\cal A}\cup {\cal A}^c$.
\end{enumerate}

For these situations it no longer suffices to compute a particular Renyi entropy for some reduced density matrix. Instead one computes the Renyi negativities for the density matrix 
$\rho_{\cal A}$, and analytically continues the even sequence of these down to $q_e \to 1$ as explained earlier. The state of the art is the computations of \cite{Calabrese:2012nk} in $d=2$ CFTs for certain specific configurations. For  instance, for ${\cal A} \subset {\mathbb R}$ being a segment of length $\ell$ bipartitioned into two segments of length $\ell\, \alpha$ and $\ell\, (1-\alpha)$ respectively the negativity was found to be
$\Eneg(\rho_{\cal A}) = \frac{c}{4}\, \log \left[\alpha\,(1-\alpha)\, \frac{\ell}{\epsilon}\right]$. The computation was made possible by explicit computation of twist operator correlation functions in $d=2$. We refer the reader to \cite{Calabrese:2012nk} for a discussion of other configurations and corresponding results for finite systems, disjoint regions, etc.. 

It should be possible to carry out in some specific  holographic situations a direct computation of the relevant quantities. We postpone this to the future, concentrating at present on the general lessons to be learnt from holography.  

\paragraph{Bipartitioning of ${\cal A}$ and phase transitions?:}
Let us start with cases (a) and (b) described above where ${\cal A}$ is partitioned into ${\cal A}_L \cup {\cal A}_R $ (case (c) is elaborated upon in \S\ref{sec:discuss}). In such cases one commonly considers the mutual information $I({\cal A}_R, {\cal A}_L)$. This is defined in terms of the entanglement entropy for the reduced density matrices induced on the two components:
\begin{equation}
I( {\cal A}_L , {\cal A}_R)  = S(\rho_{{\cal A}_L}) + S(\rho_{{\cal A}_R}) - S(\rho_{\cal A})\,.
\label{}
\end{equation}	
 If $\partial{\cal A}_L \cap \partial {\cal A}_R \neq \emptyset$ as in case (a), then both the mutual information and the negativity diverge as the area of this common boundary owing to the UV degrees of freedom in its vicinity.

There is an interesting phenomenon that occurs for holographic theories\footnote{ A necessary condition in field theory terms is that the field theories have large central charge 
$c\gg1$ (so as to admit a planar limit) and a low density of states for energies  below a gap set by $c$.\label{f:holodef}}  
in case (b) where ${\cal A}$ is composed of two disjoint regions. 
The mutual information vanishes to leading order in $c_\text{eff}$ when the regions are widely separated \cite{Ryu:2006ef}. In the holographic construction this occurs because one has to pick the globally minimal area surface (subject to boundary conditions and the topological homology constraint), which allows for phase transitions. 

Moreover, this behaviour is well understood in $d=2$ in large $c_\text{eff} = c$ CFTs in terms of a phase transition in Renyi entropies for widely separated intervals \cite{Headrick:2010zt,Hartman:2013mia}. To understand this let us describe the region by its end-points as  ${\cal A} = [u_1,v_1] \cup [u_2,v_2] \subset {\mathbb R}$. The computation of the Renyi entropy $S^{(q)}$ involves computing the four-point correlation function of ${\mathbb Z}_q$ twist operators  ${\cal T}_q$ 
\begin{align}
S^{(q)}:  \;\; \vev{{\cal T}_q(u_1) \, {\bar {\cal T}}_q(v_1) \, {\cal T}_q(u_2)\, {\bar {\cal T}}_q(v_2)} 
\label{d2ren}
\end{align}
which depends only on the cross-ratio $x = \frac{(v_1-u_1)\,(v_2-u_2)}{(u_2-u_1) \,(v_2-v_1)} \in [0,1]$ 
(up to some universal scale invariant factor). At large central charge $c$ this correlation function undergoes a phase transition at $x=\frac{1}{2}$. This is seen by decomposing the above using the OPE expansion and evaluating the contributions of the conformal block in a saddle point approximation (valid for large $c$). For small $x$ the result is dominated by the $s$-channel factorization but for $x>\frac{1}{2}$ the $t$-channel factorization takes over. In the bulk the transition is between a single connected surface and two disconnected surfaces computing $S({\cal A})$.

One might anticipate that a similar behaviour will pertain in the negativity as well since to compute the negativity one instead evaluates 
\cite{Calabrese:2012nk}
\begin{align}
\Eneg: \qquad \vev{{\cal T}_{q_e}(u_1) \, {\bar {\cal T}}_{q_e}(v_1) \, {\bar {\cal T}}_{q_e}(u_2)\, {\cal T}_{q_e}(v_2)} 
\label{d2neg}
\end{align}
Up to a switch of the insertions $u_2 \leftrightarrow v_2$ the computation is very similar to the one required for Renyi \eqref{d2ren}. The correlator \eqref{d2neg} has a non-trivial dependence on the cross-ratio $x$, in addition to  some universal contribution arising from scale invariance.  This seems to suggest that there ought be a similar phase transition in the negativity at $x=\frac{1}{2}$ for large central charge theories. 

The argument however appears to be a bit more subtle than suggested above.\footnote{ We thank Tom Hartman and Alex Maloney for discussions on this issue.} To see the issue first consider  the four-point functions relevant for the Renyi computation \eqref{d2ren}. By a suitable conformal transformation we map this to
\begin{equation}
 \vev{{\cal T}_q(0)\, {\bar {\cal T}}_q(x) \, {\cal T}_q(1)\, {\bar {\cal T}}_q(\infty) }  \equiv {\tt F}_q(x) 
\label{}
\end{equation}	
and we recall that  ${\cal T}_q$ is a twist or anti-twist operator with dimensions 
\begin{equation}
h_q = {\bar h}_q=  \frac{c}{24} \,\left(q-\frac{1}{q}\right).  
\label{}
\end{equation}	
It is sufficient to understand the behaviour of this function, since one can by utilizing the swap $u_2 \leftrightarrow v_2$ map the four-point function required for the negativity 
\eqref{d2neg} to above.  Tracking through the conformal transformations one finds  
\cite{Calabrese:2012nk} 
\begin{equation}
\frac{\vev{{\cal T}_{q}(u_1) \, {\bar {\cal T}}_{q}(v_1) \, {\bar {\cal T}}_{q}(u_2)\, {\cal T}_{q}(v_2)}}{\vev{{\cal T}_q(u_1) \, {\bar {\cal T}}_q(v_1) \, {\cal T}_q(u_2)\, {\bar {\cal T}}_q(v_2)} }
= (1-x)^{8\,h_q}\, \frac{{\tt F}_q \left(\frac{x}{x-1}\right) }{ {\tt F}_q(x)}
\label{ratioSE}
\end{equation}	

We thus have a direct link between the two computations and all we need is the behaviour of the function ${\tt F}_q(x)$. One has control on this function for $x \in [0,1]$ from the analysis of 
\cite{Hartman:2013mia} in the large $c$ limit (cf.,  footnote \ref{f:holodef}), which can be used to argue that the Renyi entropies undergo a phase transition. To make an argument for the negativity however requires that we also control the function outside this domain. It is tempting to conjecture that the phase transition does indeed happen and moreover one encounters a similar behaviour in higher dimensions. We leave a more detailed analysis for the future.

\section{Discussion}
\label{sec:discuss}

 In this paper we have focussed on properties of entanglement negativity, defined as a measure of distillable entanglement in a given state of a quantum system. The rationale for its definition lies in understanding the entanglement structure of mixed states. To gain some intuition for this quantity we explored its properties in simple states such as the vacuum of a CFT in various dimensions. While we laid out some general expectations for the behaviour of negativity in holographic field theories more generally, we did not offer any concrete computations in supporting evidence. We hope to remedy this in the near future.  
 It is nevertheless useful to take stock and examine some of the questions posed by the analysis we have undertaken.

First of all, it is interesting to ask if there is some intrinsic meaning to the geometric pre-factor ${\cal X}_{\cal A}$.  Since $\Eneg$ provides only an upper bound on the distillable entanglement,  what physical interpretation, if any, should be ascribed to its being greater than the entanglement entropy? Can one think of ${\cal X}_{\cal A}\, c_\text{eff}$ as a measure of the  effective number of Bell pairs that can be distilled out of a pure state in a CFT? 
 
 We have also seen earlier that this function renormalizes and for spherically symmetric regions ${\cal X}_{\cal A}$ it was smaller (in magnitude) at strong coupling. Does this reduced amount in distillable entanglement have a fundamental significance in how spacetime geometry is related to the presence of entanglement? It would be instructive to know whether one can formalize some statement along these lines in a quantitative fashion. At a more prosaic level it would be interesting to understand this function both as a function of the state $\psi$ as well as the geometry of the region ${\cal A}$. 

Secondly, all of our discussion has been restricted to density matrices at a moment of time symmetry (or in special cases static density matrices). This allowed us in the general context to make use of the generalized gravitational entropy construction of \cite{Lewkowycz:2013nqa} to compute the Renyi entropies and negativities for integer values of the index $q$. These are clearly special situations and one would like to be able to make statement for time-evolving states. As in the case of entanglement entropy extending the construction to dynamical situations could perhaps teach us some new lessons about spacetime and entanglement.

As a final comment, we turn to the situation where ${\cal A}$ is a single connected region, but one has a mixed state on the entire Cauchy slice $\Sigma$ (denoted $\rho_\Sigma$) (case (c) in \S\ref{sec:mixed}).  As remarked earlier one of the main reasons to focus on negativity is to understand the precise nature of entanglement in mixed states. In the holographic context one encounters an interesting feature for the entanglement entropy of reduced density matrices $\rho_{\cal A}$ induced from a parent thermal state.  When ${\cal A}$ is a sufficiently large region of the Cauchy slice one finds an interesting phenomena dubbed entanglement plateaux \cite{Hubeny:2013gta}: $S(\rho_{\cal A}) = S(\rho_{{\cal A}^c}) + S_{\rho_{\Sigma}}$, i.e., Araki-Lieb inequality \cite{Araki:1970ba} is saturated. This behaviour has been argued to be robust in holographic field theories for finite systems at large $c_\text{eff}$.

One can interpret this to mean that the entanglement  inherent in $\rho_{\cal A}$ has two distinct contributions: (i) the quantum entanglement between the region and its complement  encapsulated in $S(\rho_{{\cal A}^c}) $ and (ii)  correlations built into the thermal density matrix $S_{\rho_\Sigma}$. This distinction seems to suggest that in this regime there is a clear demarcation in the degrees of freedom inside ${\cal A}$ in terms of their entanglement properties  \cite{Headrick:2013zda} (see also \cite{Zhang:2012fp} for related considerations). Indeed this interpretation is natural from the perspective of the disentangling theorem for tripartite systems described in 
\S\ref{sec:negative}. The thermofield double state which purifies the density matrix $\rho_\Sigma$ factorizes as in \eqref{trineg1} with $B = {\cal H}_{\cal A}$.  It would be fascinating to see this arise directly by computing the negativities in the holographic context.

\acknowledgments 

It is a pleasure to thank Horacio Casini, Felix Haehl, Tom Hartman, Veronika Hubeny, Henry Maxfield and Tadashi Takayanagi for useful discussions on various aspects of quantum entanglement. We would also like to thank Horacio Casini, Veronika Hubeny, Tadashi Takayanagi and Erik Tonni for comments on a preliminary draft of the paper.
M.~Rangamani would like to thank the Yukawa Institute for Theoretical Physics, Kyoto for hospitality during the concluding stages of this project. M.~Rangamani was supported in part by FQXi  grant "Measures of Holographic Information" (FQXi-RFP3-1334),  by the STFC Consolidated Grant ST/J000426/1 and by the European Research Council under the European Union's Seventh Framework Programme (FP7/2007-2013), ERC Consolidator Grant Agreement ERC-2013-CoG-615443: SPiN (Symmetry Principles in Nature).


\providecommand{\href}[2]{#2}\begingroup\raggedright\endgroup

\end{document}